\documentclass{czjphys}         
\usepackage{epsfig}  
\begin{document}
\title{Models for cosmic ray interactions}  
%
\authori{S S Ostapchenko}      
\addressi{Forschungszentrum Karlsruhe, Institut f\"ur Kernphysik,
            76021 Karlsruhe, Germany\\
	    D.V. Skobeltsyn Institute of Nuclear Physics,
         Moscow State University,\\ 119992 Moscow, Russia}
\authorii{}
\addressii{}
\authoriii{}    \addressiii{}
\authoriv{}     \addressiv{}
\authorv{}      \addressv{}
\authorvi{}     \addressvi{}
%
\headauthor{S S Ostapchenko}            
\headtitle{Models for cosmic ray interactions}             
\lastevenhead{S S Ostapchenko} 
\refnum{A}
\daterec{XXX}    
\issuenumber{0}  \year{2001}
\setcounter{page}{1}
\maketitle
\begin{abstract}
 Contemporary models of hadronic interactions are reviewed. 
 Basic phenomenological approaches are compared, with an 
 emphasizes on the predicted air shower characteristics. 
 Special attention is payed to the remaining discrepancies 
 between present hadronic MC generators and cosmic ray data. 
 Finally, future prospects concerning model improvements are
  discussed, in particular, regarding the possibilities to
   discriminate between different models on the basis of
    accelerator or cosmic ray measurements. 
\end{abstract}

\section{Introduction}
Nowadays Monte Carlo (MC) models of hadronic interactions are extensively
used in collider and  cosmic ray (CR) physics, being applied for 
designing new experiments, analyzing and interpreting data, or for testing
new theoretical ideas, when the latter are incorporated in the
corresponding MC generators. Here the last option is more characteristic
for accelerator physics. High energy cosmic rays are typically studied
by means of extensive air shower (EAS) techniques: instead of detecting
primary CR particles directly one measures different characteristics
of the nuclear-electro-magnetic cascade, so-called air shower, induced by
their interactions in the atmosphere. Because of a large number of 
hadronic collisions  during such a cascade 
it is very difficult to disentangle  partial contributions of 
interactions of  particular energies and to test individual features of
MC models. In CR physics hadronic generators are an important technical tool,
which  extrapolates
current theoretical and experimental knowledge from the energies of present
accelerators up to the highest CR energies and helps to establish
a relation between the properties of primary cosmic ray particles and 
the observed EAS characteristics.

Are there significant differences between collider and cosmic ray models?
 Is it possible to apply popular
accelerator codes for CR data analysis? As we shall discuss below,
all contemporary hadronic MC generators are based on similar
physics ideas. Both at colliders and in cosmic rays one employs
various approaches to describe  hadronic and nuclear interactions;
in neither case there exist a preferred approach, inherent for this or that
particular field. However, generally accelerator codes are not
suitable for an immediate application in CR field, being unable
 to treat some necessary reactions,
like interactions of pions and kaons with nuclei or
 minimum bias nucleus-nucleus collisions,
  to perform predictive calculations of total inelastic
and diffractive cross sections, or to  be extrapolated
 into the ultra-high energy domain. It is worth stressing again that cosmic
 rays experiments can  test model validity only indirectly. Analyzing
 CR data one has to rely on model predictions, without real possibilities
 to re-tune model parameters or to refine corresponding algorithms.

  The goal of this work is to review  basic approaches employed in
 contemporary CR  models,
to analyze remaining problems and to outline promising directions for
future model  improvements and for
 model  discrimination  on the basis of
    accelerator and cosmic ray measurements.

\section{High energy interactions: ``pure QCD'' approach}
Nowadays perturbative quantum chromo-dynamics (pQCD) is the only strict
 theory which provides the basis for both
qualitative and quantitative description of hadronic production 
in high energy reactions. However, precise pQCD calculations are only
possible for a restricted class of processes, namely those which are
characterized by a large momentum transfer $Q^2$, such that running
 coupling $\alpha _s(Q^2)$ becomes small
and corresponding contributions can be re-summed to a reasonable accuracy
 \cite{alt82}.
Thus, one can not  apply it to general hadron-hadron processes, which are
expected to be dominated by the ``soft'' (low $Q^2$) physics.
On the other hand, in the very high energy limit one expects that the role
of high virtuality (high $p^2_t$) processes greatly increases, as their
contributions are enhanced both by large parton multiplicity and by large
logarithmic  ratios of transverse and longitudinal momenta for successive
parton emissions \cite{alt82,glr}. Hence, one may develop MC generators,
 which rely on the pQCD formalism in the description
of ``hard''    (high $p_t$) processes while treating the ``soft'' ones
in a simplified fashion, assuming that the latter play smaller and smaller
role when the energy increases.

Quite often such generators are referred to as ``pure QCD models''.
How much is defined there by the perturbative QCD? Typically the main
pQCD input  is the inclusive cross section for  production
of parton jet pairs with transverse momenta larger than some cutoff  $Q_0$,
given to  leading logarithmic accuracy as 
\begin{eqnarray}
\sigma ^{\rm {jet}}_{ad}\! \left(s,Q_{0}^{2}\right) &=&
\sum _{i,j=g,q,\bar q}\int \! \frac{dp^{2}_t}{p^{2}_t}
\int \! \frac{dx^{+}_{i}dx^{-}_{j}}{x^{+}_{i}x_{j}^{-}}\, 
\frac{d\sigma ^{2\rightarrow 2}_{ij}\left( x_{i}^{+}x_{j}^{-}s
,p^{2}_t\right)}{dp^{2}_t}   \label{sigjet} \\
&\times & f^{i}_{a}\left( x_{i}^{+},p^{2}_t\right)
 \; f^{j}_{d}\left( x_{j}^{-},p^{2}_t\right)
  \; \Theta \left( p^{2}_t-Q_{0}^{2}\right)\,, \nonumber 
\end{eqnarray}
where  $s$ is the c.m. energy squared for the interaction,
$d\sigma ^{2\rightarrow 2}_{ij}/dp^{2}_t$  is the lowest
order  parton-parton scattering cross section, and
$f^{i}_{a}\left( x,q^{2}\right)$ is  the parton $i$ ((anti-)quark or gluon)
 momentum
 distribution function (PDF) in hadron $a$, when probed at the virtuality 
 scale $q^2$. Here partons $i$ and $j$ in Eq.~(\ref{sigjet}),  participating
 in the high $p_t$ jet production,
  originate from partonic cascades in hadrons $a$ and $d$,
 being emitted by parent partons of smaller $p_t$ and larger energy, those
 in turn being emitted by their own parents and so on. The whole process
 can be described as a parton ladder, which starts in hadrons  $a$ and $d$
 with partons of  $p_t^2\simeq Q_0^2$.
 
\begin{figure}[htb]
\begin{center} 
\includegraphics[width=7cm,height=2.5cm]{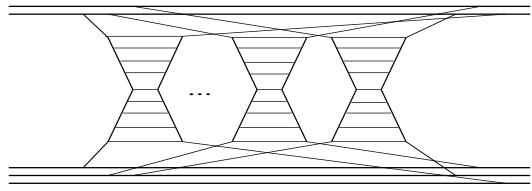}
\vspace*{-.7cm}
\end{center} 
\caption{Hadron-hadron scattering as multiple exchange of parton ladders.
\label{mult-lad}} 
\end{figure} 
 Still, to develop a MC model one has to make a number of additional 
 assumptions, which are the essence of the mini-jet approach
 \cite{dur87,gai89} employed in  SIBYLL  \cite{fle94,eng99}
  and in a number of 
 popular collider models, like PYTHIA \cite{sjo87} or HIJING \cite{wan97}.
 First, as  $\sigma ^{\rm {jet}}_{ad}\! \left(s,Q_{0}^{2}\right)$ 
 increases with energy faster than the observed total cross section, 
 it appears mandatory to consider
 multiple jet production processes, in other words, to describe the
  interaction as  multiple exchange of parton ladders, see 
  Fig.~\ref{mult-lad}. 
  
  Furthermore, one suggests that parton  distribution
  in  hadron $a$ is governed by its electro-magnetic form factor 
  $\rho _a^{\rm e/m}$  and,
  as a consequence, that at  given impact parameter $b$
   average number of jet production processes
  is defined by the product of $\sigma ^{\rm {jet}}_{ad}$ and the hadron-hadron
  overlap function 
  $A_{ad}(b)=\int \!d^2b'\: \rho _a^{\rm e/m}(b')\:
  \rho _d^{\rm e/m}(|\vec b -\vec b'|)$:
  \begin{equation}
  \langle n ^{\rm {jet}}_{ad}\!\left(s,b,Q_{0}^{2}\right)\rangle
  =\sigma ^{\rm {jet}}_{ad}\! \left(s,Q_{0}^{2}\right)\;
  A_{ad}(b)
  \end{equation}
 
 Finally, assuming Poisson distribution for
  the number of jet processes at given  $b$, with the average
 value $ \langle n ^{\rm {jet}}_{ad}\!\left(s,b,Q_{0}^{2}\right)\rangle
 \equiv 2\chi^{\rm {hard}}_{ad}\!\left(s,b\right)$, and adding a similarly
 defined contribution of ``soft'' processes 
 $\chi^{\rm {soft}}_{ad}\!\left(s,b\right)
 =\frac 12 \sigma^{\rm {soft}}(s)\:A_{ad}(b)$, with a parameterized
 bare ``soft'' cross section $\sigma^{\rm {soft}}(s)$, one can express
  total inelastic cross section via the  eikonal
  $\chi _{ad}=\chi^{\rm {hard}}_{ad}+\chi^{\rm {soft}}_{ad}$ 
  as \cite{gai89,wan97}
 \begin{equation}  \label{sigs}
\sigma ^{\rm inel}_{ad}(s)= \int \!d^{2}b\,
\left[1-e^{-2\chi _{ad} (s,b)}\right] \label{sigm-inel}
\end{equation}

Using this scheme one can  generate inelastic events,
 starting from sampling the impact
parameter $b$ for the interaction, defining the number of  hard
processes, and 
generating parton cascades which preceed the hardest parton-parton
scattering or which follow after it \cite{sjo87,wan97}. Then, one
employs phenomenological procedures to convert  final partons
into hadrons, as well as to treat the soft part of the interaction.
The most popular one is the string picture \cite{ben87}:
 when final partons move apart,
being connected to each other by color field, a color string is stretched 
between them. As long as the distance between partons increases so does the
string tension, which finally gives rise to the break up of the string 
 and to a multiple 
creation of quark-antiquark and diquark-antidiquark pairs,
those forming secondary hadrons.

The disadvantage of the approach is a somewhat artificial separation of
 hard and soft processes and a rather arbitrary treatment for the latter.
 For the former one considers parton cascades as starting at  given
 cutoff scale $Q_0^2$, thus neglecting the contribution of low $p_t$ 
 ($p_t<Q_0$) partons.

\section{High energy interactions: Reggeon approach}
\begin{figure}[htb]
\begin{center} 
\includegraphics[width=7cm,height=2.5cm]{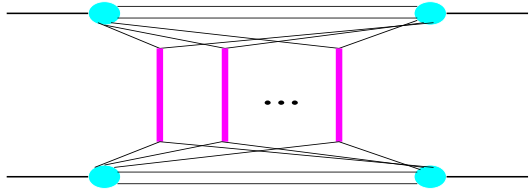}
\vspace*{-.7cm}
\end{center} 
\caption{Typical contribution to hadron-hadron scattering amplitude;
 elementary re-scatterings are  treated as exchanges of
composite objects -- Pomerons.\label{mult}} 
\end{figure}

In  Gribov's Reggeon approach \cite{gri68} hadronic 
 collisions are described as multiple scattering processes, where
each elementary re-scattering corresponds to a  microscopic  parton cascade,
 just like in Fig.~\ref{mult-lad}.
Such an elementary  process is treated there
phenomenologically, as an exchange of a quasi-particle, the
 Pomeron, so that the scattering amplitude  is the sum
 of contributions of graphs of Fig.~\ref{mult}.

The Pomeron  amplitude is postulated in the form \cite{kai84}: 
\begin{eqnarray}   \label{softpom}
 f^{\rm P}_{ad}\! \left(s,b\right) & =&
  i\gamma_{a}\gamma_{d}\,s^{\alpha_{\rm P}(0)-1}/\lambda_{ad}(s)\,
 \exp \!\left[ -b^{2} /\left(4 \lambda_{ad}(s)\right)\right]\equiv 
  i\chi^{\rm P}_{ad}\! \left(s,b\right) \\
  \lambda_{ad}(s)&=  & R_{a}^{2}+R_{d}^{2}
  + \alpha_{\rm P}'(0)\ln s, 
\end{eqnarray}
which is characterized by a power-like energy rise and  by a logarithmically
increasing slope $\lambda_{ad}(s)$. Here
$\alpha_{\rm P}(0)$, $\alpha_{\rm P}'(0)$ are the Pomeron trajectory
 parameters (intercept and slope) and $\gamma _a$, $R^2_a$ are
 the parameters for  Pomeron-hadron $a$ coupling.
 
Having specified the structure of Pomeron-hadron vertices, 
one can obtain the total elastic amplitude for hadron-hadron scattering,
hence, via the optical theorem, also total cross section.
Applying the Abramovskii-Gribov-Kancheli (AGK) cutting rules 
  \cite{agk}, one can separate  different contributions to the total
  cross section, corresponding to particular final channels of the reaction.
 For example, assuming eikonal vertices and neglecting diffraction
  dissociation and inelastic screening, one  obtains
   Poisson distribution for the number of elementary
  production processes at given $b$ and arrives to  Eq.~(\ref{sigm-inel})
  for  
 inelastic cross section, with 
   $\chi _{ad}(s,b)=\chi^{\rm P}_{ad}(s,b)$.

\begin{figure}[htb]
\begin{center}
\includegraphics[width=8cm,height=3cm]{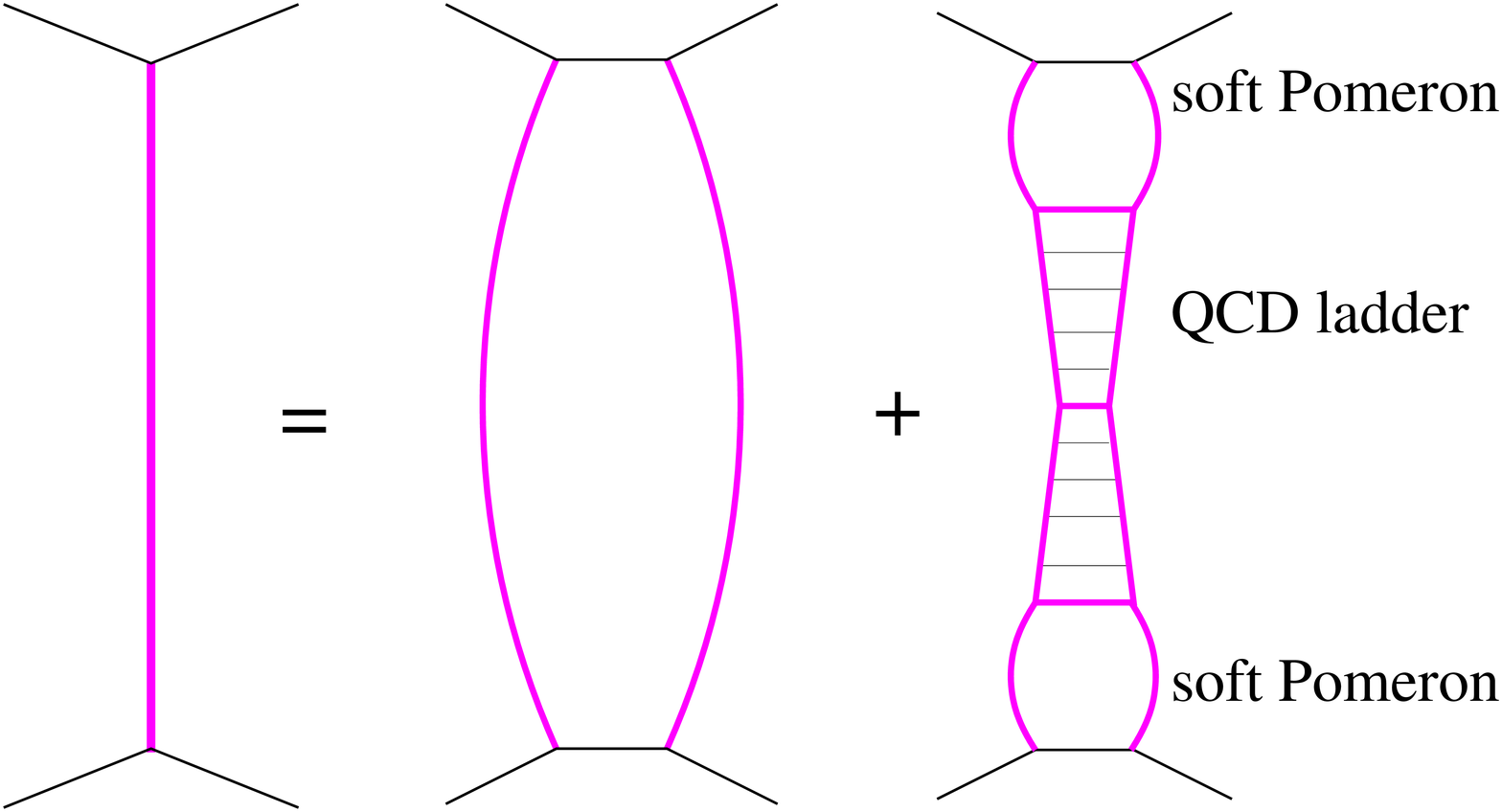}
\vspace*{-.7cm}
  \end{center}
\caption{A general ``Pomeron'' (l.h.s.) is the sum  of the soft and
the  semi-hard ones -- correspondingly the 1st and the 2nd
graph in the r.h.s. \label{genpom} }
\end{figure}

How to match this scheme with the pQCD treatment?
 In general,   parton cascades contain both ``hard''
 ($p_t^2 >Q^2_0$) and ``soft'' ($p_t^2 <Q^2_0$) parts,
 $Q_0$ being a chosen cutoff for pQCD being applicable.
 In the ``semi-hard Pomeron'' scheme, employed in QGSJET \cite{kal94} and 
 NEXUS  \cite{dre01} models,
 one applies phenomenological soft Pomeron treatment for the soft
 part of the parton cascade and describes general  semi-hard processes
  as exchanges of a ``semi-hard Pomeron'', the latter being represented by 
  a piece of QCD ladder sandwiched between two  soft Pomerons 
\cite{kal94,dre99,ost02}, see Fig.~\ref{genpom}.
A general Pomeron appears to be the sum of the semi-hard and soft ones,
the latter corresponding to a cascade of low $p_t$
($p_t <Q_0$) partons.

The scheme coincides with the mini-jet approach in the treatment of
the hard sector, represented by the ladder parts of the semi-hard Pomerons,
see Fig.~\ref{genpom}.
An important difference comes from considering multiple soft interactions,
described as soft Pomeron exchanges, and from accounting for an additional
source of particle production, which comes from the low $p_t$ part of the
parton cascades (``soft pre-evolution'') in semi-hard processes,
shown as soft Pomeron ``blobs'' in Fig.~\ref{genpom}. 
There is no artificial separation between soft and semi-hard processes,
both being  partial contributions to a general parton cascade,
depending whether the latter develops in the low  $p_t$ region
or extends to higher virtualities, $p_t^2 >Q^2_0$. The same soft Pomeron 
asymptotics is used to describe both soft processes and  soft parts of  
semi-hard ones, defining in the latter case corresponding PDFs at  
scale $Q^2_0$ \cite{dre01,ost02}.

In general, there is no sharp border between the models of mini-jet or
semi-hard Pomeron type. For example,  modern version of  SIBYLL
 \cite{eng99} takes into account multiple soft interactions,
described as  Pomeron exchanges; DPMJET \cite{ran95} considers an
 additional soft Pomeron exchange for each semi-hard process, which is not
 very different from the above-described ``soft pre-evolution'' treatment. 

\section{Non-linear effects in high energy collisions}\label{nonlin}
\begin{figure}[htb]
\begin{center} 
\includegraphics[width=5cm,height=2.5cm]{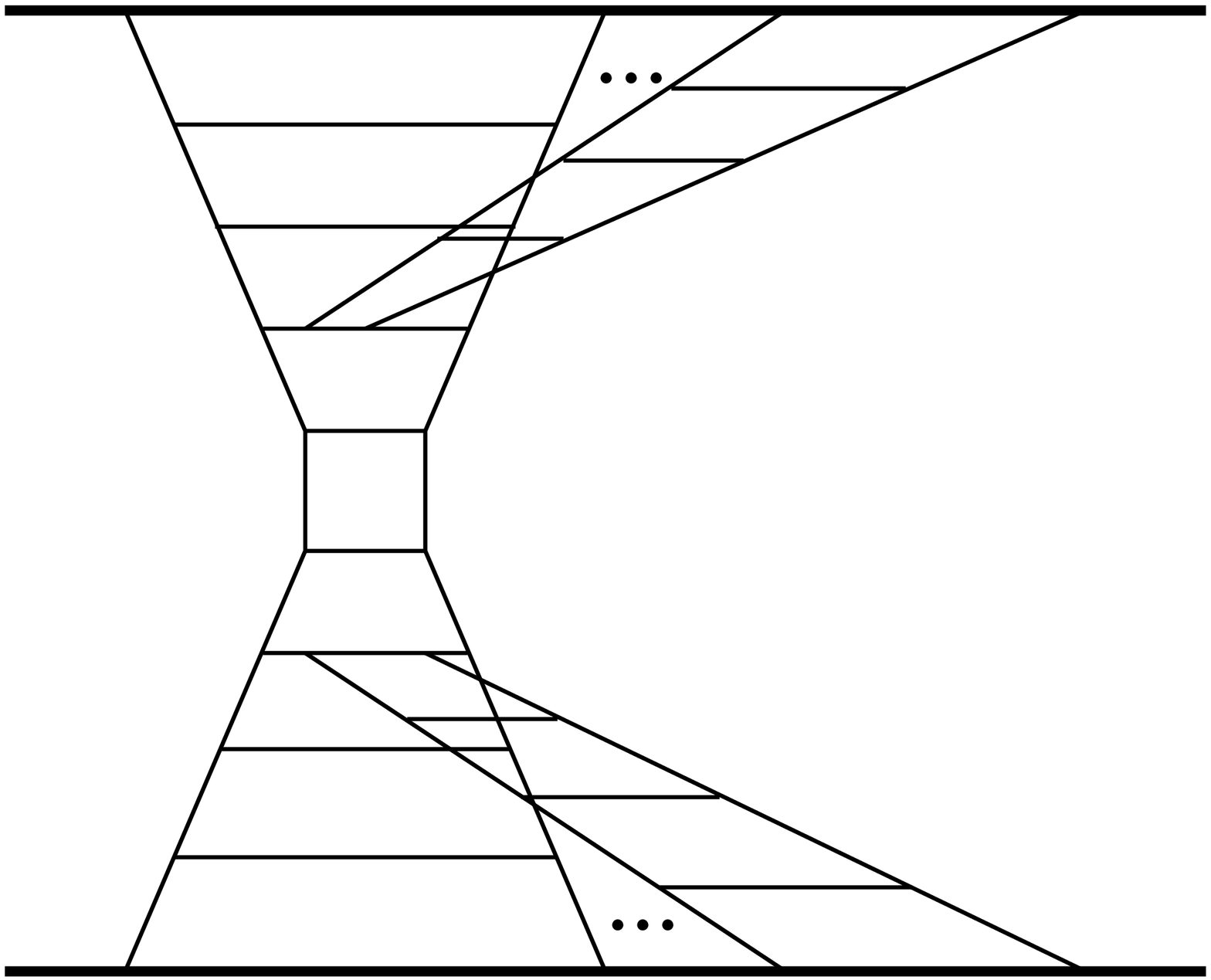}
   \hspace{1cm}
\includegraphics[width=5cm,height=2.5cm]{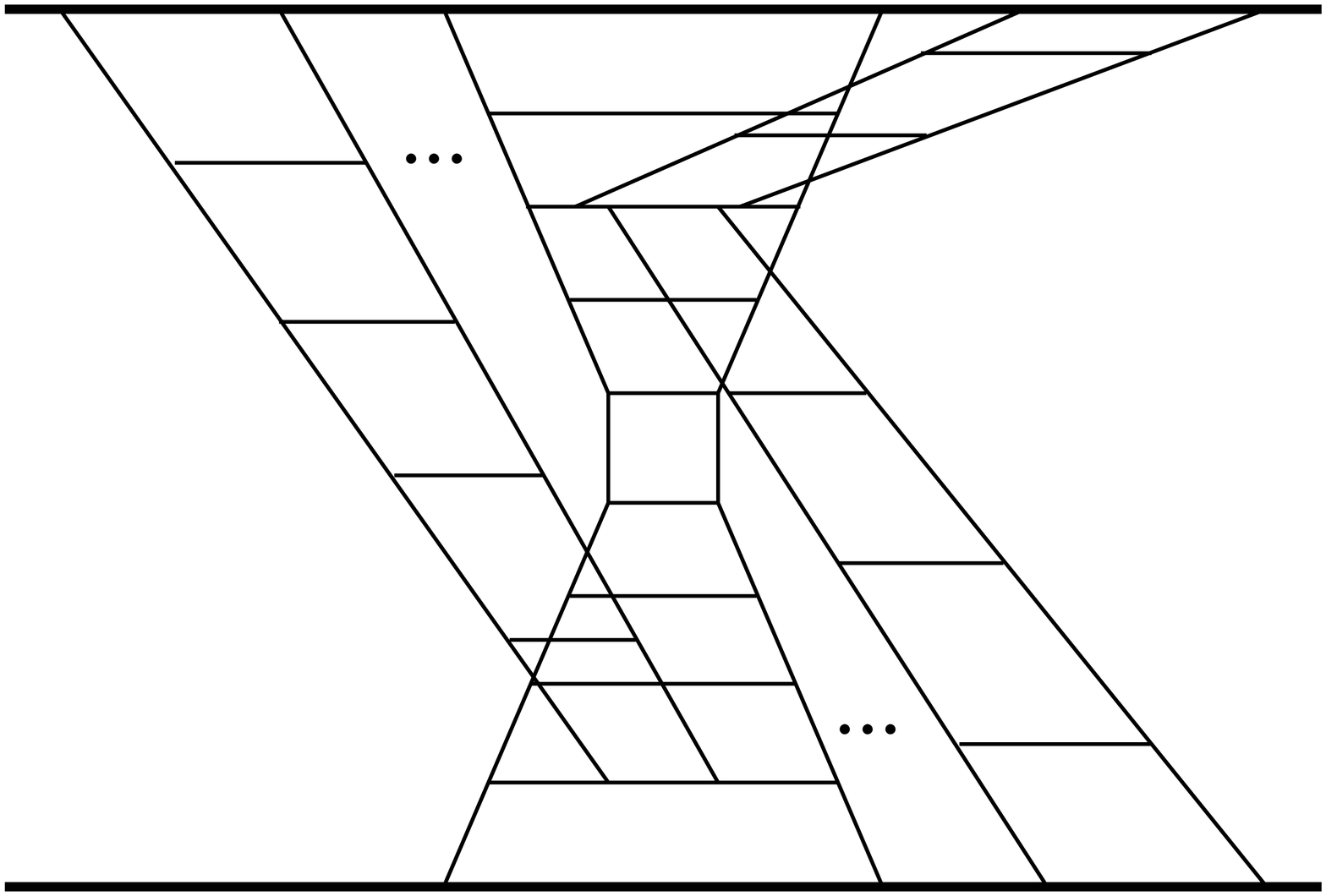}
\vspace*{-.7cm}
\end{center} 
\caption{Examples of diagrams which give rise to non-linear parton effects.
   \label{non-lin}} 
\end{figure}

Treating interactions at very high energies and small impact
parameters one deals with  the regime of high parton densities. There,
individual parton cascades are no longer independent of each other;
their overlap and mutual influence give rise to  significant
modifications of the picture described so far. At microscopic level
such non-linear  effects are described as merging of parton ladders
\cite{glr}, see Fig.~\ref{non-lin}. 

The corresponding pQCD treatment is 
 based on the assumption that parton densities in the 
low virtuality region are saturated and any further parton branchings 
are compensated  by the fusion of individual parton cascades \cite{glr}.
 In mini-jet models one typically
employs a phenomenological approach, introducing some parameterized
energy-dependence of the $p_t$-cutoff for mini-jet production, 
$Q_0=Q_0(s)$, the latter being
 an effective saturation scale \cite{eng99}.
 Still, completely neglecting the contribution of the saturated region
 (partons of $p_t<Q_0(s)$) to  particle production
  may be a too strong assumption. On the other hand, applying the same
  $Q_0(s)$ cutoff to all collisions at a given energy appears to be a
  rather crude approximation; one loses   correlations with actual
  parton densities, which depend on the parton
  Feynman $x$, on the impact parameters, and on the projectile and target
  mass numbers in case of nuclear collisions.

A dynamical description of non-linear interaction effects
has been proposed in the QGSJET-II model \cite{ost04},
based on the assumption that saturation effects can be neglected for parton
virtualities bigger than some fixed, energy-independent cutoff $Q_0$.
In this scheme the multi-ladder
graphs of Fig.~\ref{non-lin} are replaced by  enhanced 
diagrams, which correspond to Pomeron-Pomeron interactions \cite{kan73,kai86}.
 Suggesting that multi-Pomeron vertices are dominated by
 parton processes at comparatively low virtualities $|q^2|<Q_0^2$,
one can re-sum all significant contributions of that kind and  develop
 a self-consistent MC generation procedure for hadronic and nuclear collisions
\cite{ost04}. The main parameter of the scheme, the triple-Pomeron
coupling, has been inferred from HERA data on hard diffraction
in deep inelastic scattering reactions. 
By construction,  non-linear screening corrections appear to be
 correlated with corresponding parton densities and become larger
at higher energies, smaller impact parameters, or for collisions of heavier
nuclei, resulting  finally  in the saturation of PDFs at the scale $Q_0^2$
and in a considerable reduction of ``soft'' particle production.
In particular,  the predictions of the QGSJET-II model stay in agreement 
 with the data of the RHIC collider on the
multiplicity of secondary hadrons produced in central nucleus-nucleus
interactions \cite{ost06}.

\section{Model sensitivity of air shower calculations}

Applying different models to air shower calculations one typically obtains
some spread of the predicted EAS characteristics. To get insight into
 such differences, one compares different models with
 respect to the interaction characteristics which are most relevant 
 for air shower development, i.e.~regarding
the predicted energy rise of inelastic cross sections,
 the relative energy loss of ``leading'' secondary particles 
 (the inelasticity),  or the multiplicity
of produced secondaries. Such  comparisons  help to understand 
 discrepancies between  calculations and experimental data and
  to outline directions for future model improvements.

On the other hand, not every problem of EAS data interpretation 
may be attributed to potential model deficiencies. 
Nowadays there exist a bulk of accelerator data
which seriously constrain  predictions of hadronic  models. 
Moreover, due to a large number of  interactions
in a hadronic cascade,
 observed air shower characteristics appear to  depend rather
on the general interaction trend over a wide energy range 
than on a sudden change 
of the interaction mechanism at a particular energy.
 In particular, comparatively 
low energy interactions, being well studied at accelerators, 
contribute with a large weight \cite{mei05}. Thus, it would be naive to expect  
that some new valid model will change basic EAS predictions by a large amount.

\begin{figure}[htb]
\begin{center} 
\includegraphics[width=5.5cm,height=4.1cm]{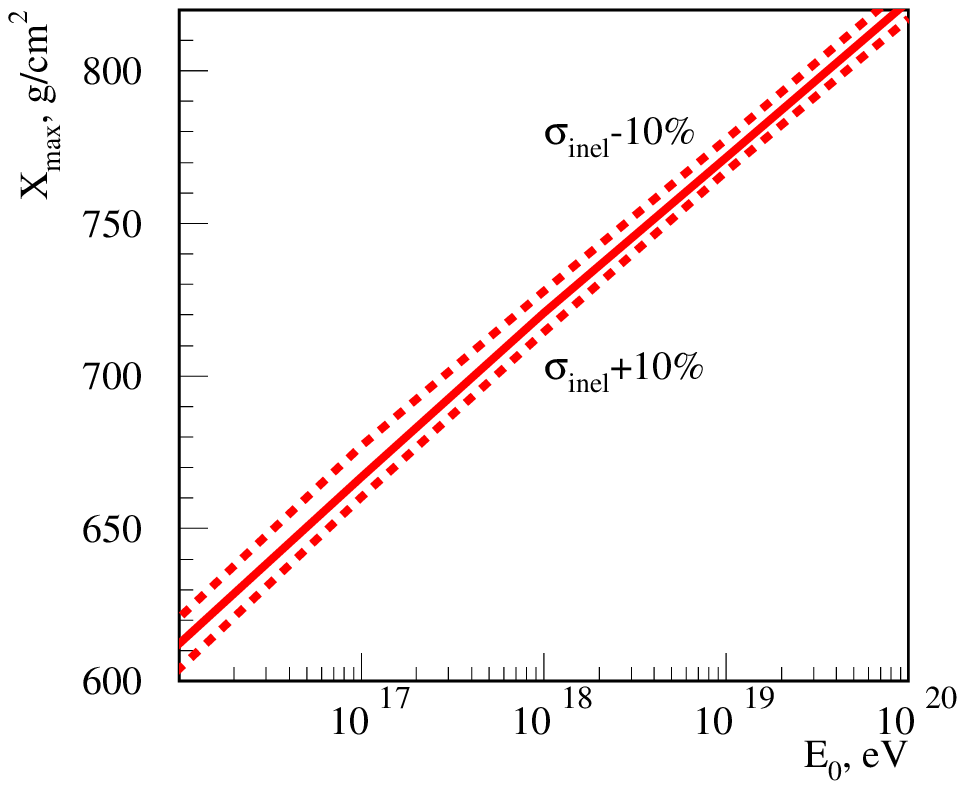}
\hspace{1cm}
\includegraphics[width=5.5cm,height=4.1cm]{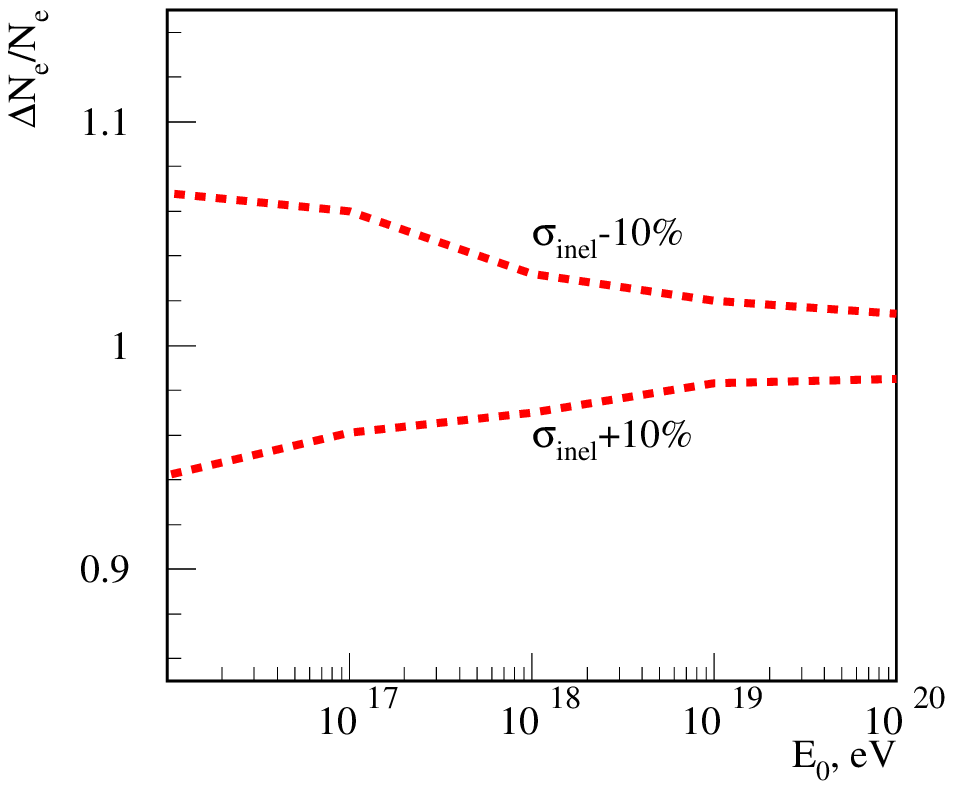}
\vspace*{-.9cm}
\end{center} 
\caption{Left: shower maximum depth for proton-induced EAS
 as calculated with the default QGSJET
model (full line) and  with $\pm 10$\% modifications of
$\sigma ^{\rm inel}_{\rm p-air}$ (dashed lines).
Right: variations of the calculated electron number at sea level
 for proton-induced vertical EAS for $\pm 10$\% modifications of
$\sigma ^{\rm inel}_{\rm p-air}$ (relative to the default QGSJET).
 \label{sig+10}} 
\end{figure}

To quantify these statements one can perform simple tests with
 presently available models.
For example, using the QGSJET model with the corresponding proton-air cross section
being  artificially rescaled by $\pm 10$\%, while keeping pion-air cross 
section unchanged,
 one obtains, due to the shift of the primary particle interaction point,
  about 10 g/cm$^2$ variation of the predicted shower 
maximum depth $X_{\max}$; the corresponding change of the charged
particle number $N_e$ at ground  ranges from 6\% at $10^{16}$ eV
to 2\% at $10^{19}\div 10^{20}$ eV,
 as shown in Fig.~\ref{sig+10}. 
For  charged particle density
at 1000 m distance from the shower core $\rho _{\rm ch}(1000)$
such a  sensitivity is even weaker,
falling below 1\% above $10^{18}$ eV, as the reduction (enhancement)
 of $N_e$ at ground is compensated by a flatter (steeper) lateral distribution
 of particles in case of larger (smaller) cross section.
Thus, present experimental techniques, which reconstruct the primary energy
on the basis of $\rho _{\rm ch}(1000)$, should not 
 exhibit any significant model dependence. A similar conclusion holds
 for the fluorescence light-based energy reconstruction procedures
  \cite{pie05}.

\begin{figure}[htb]
\begin{center} 
\includegraphics[width=5.5cm,height=3.7cm]{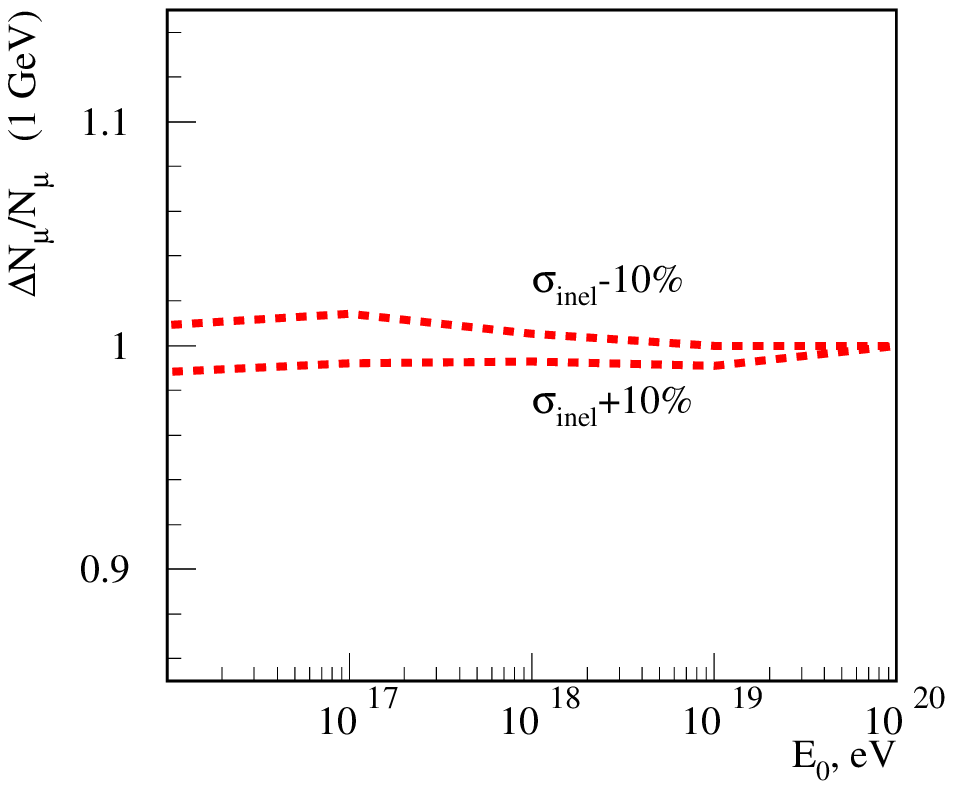}
\hspace{1cm}
\includegraphics[width=5.5cm,height=3.7cm]{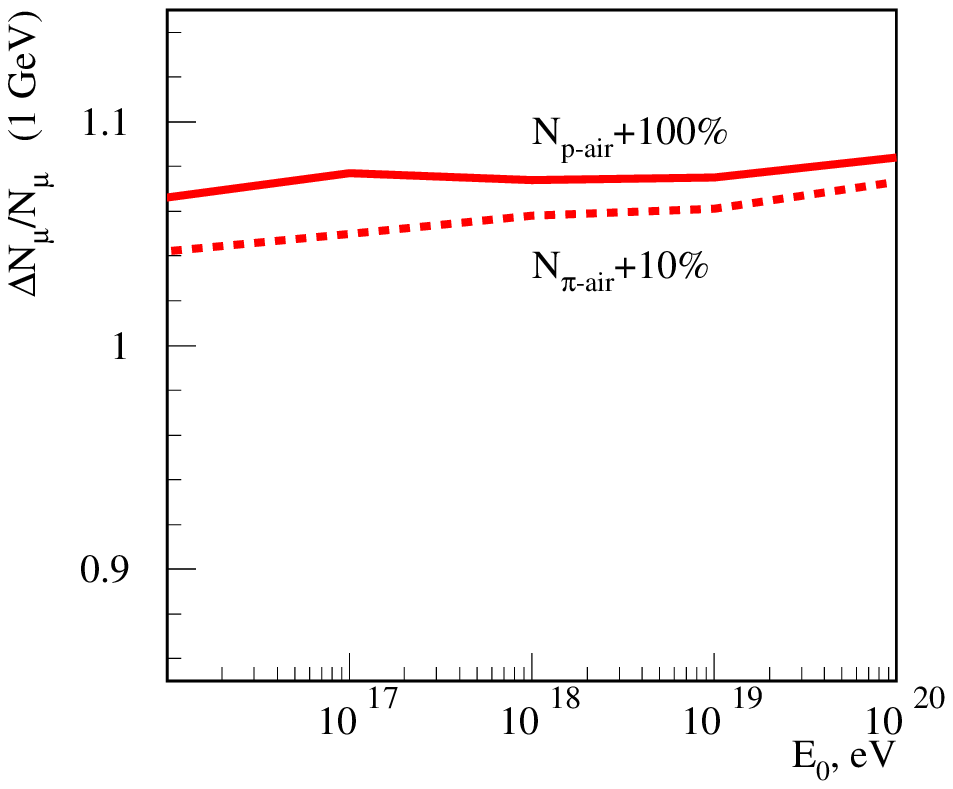}
\vspace*{-.9cm}
\end{center} 
\caption{Sensitivity of the  muon number ($E_{\mu}>1$ GeV) at sea level
for proton-induced vertical EAS to the cross section and multiplicity
variations (as discussed in the text). \label{muons}} 
\end{figure}

Still, any studies of the cosmic ray composition,
 either based on  $X_{\max}$
measurements or making use of  electron-muon number correlations, 
exhibit significant
 model sensitivities. As for the shower muon number $N_{\mu}$ at ground level,
  it is very
robust with respect to the discussed cross section variation, as shown 
in  Fig.~\ref{muons}(left),
but depends  on the multiplicity of hadronic interactions. 
One can consider
two alternative modifications of model predictions -- 
either increasing the multiplicity
of just the primary particle interaction by a factor of two or 
enhancing the multiplicity 
of all secondary pion-air and kaon-air interactions by 10\% -- 
see Fig.~\ref{muons}(right).
As is easy to see from the Figure, both changes make a similar impact on the predicted
EAS muon number at highest energies. Evidently, to obtain a factor of two
 increase of  $N_{\mu}$ 
one  needs a comparable enhancement of  multiplicities
 of both primary and all secondary interactions,
 which would strongly contradict available accelerator data.

What are  characteristic differences between  presently available 
models? QGSJET and SYBILL predictions 
for  $\sigma ^{\rm inel}_{\rm h-air}$ differ from each other
 by  $10\div 15$\%  at $10^{19}$ eV;
 the disagreement for the multiplicity of 
 charged particles $N^{\rm ch}_{\rm p-air}$
\begin{figure}[htb]
\begin{center} 
\includegraphics[width=5.5cm,height=3.7cm]{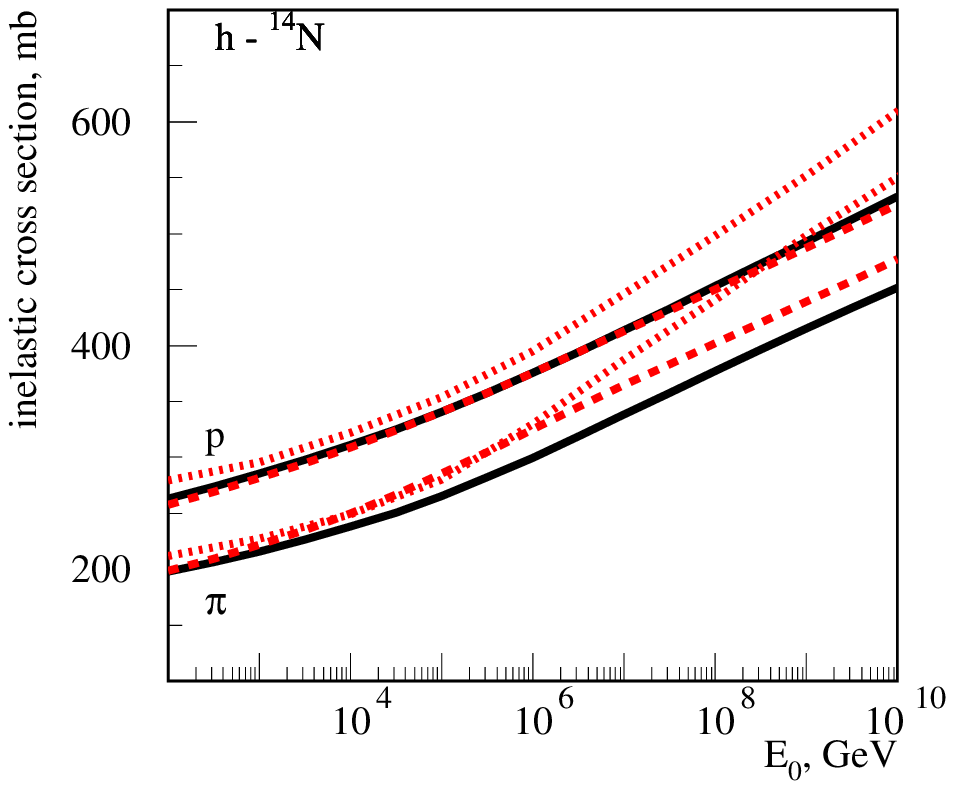}
\hspace{1cm}
\includegraphics[width=5.5cm,height=3.7cm]{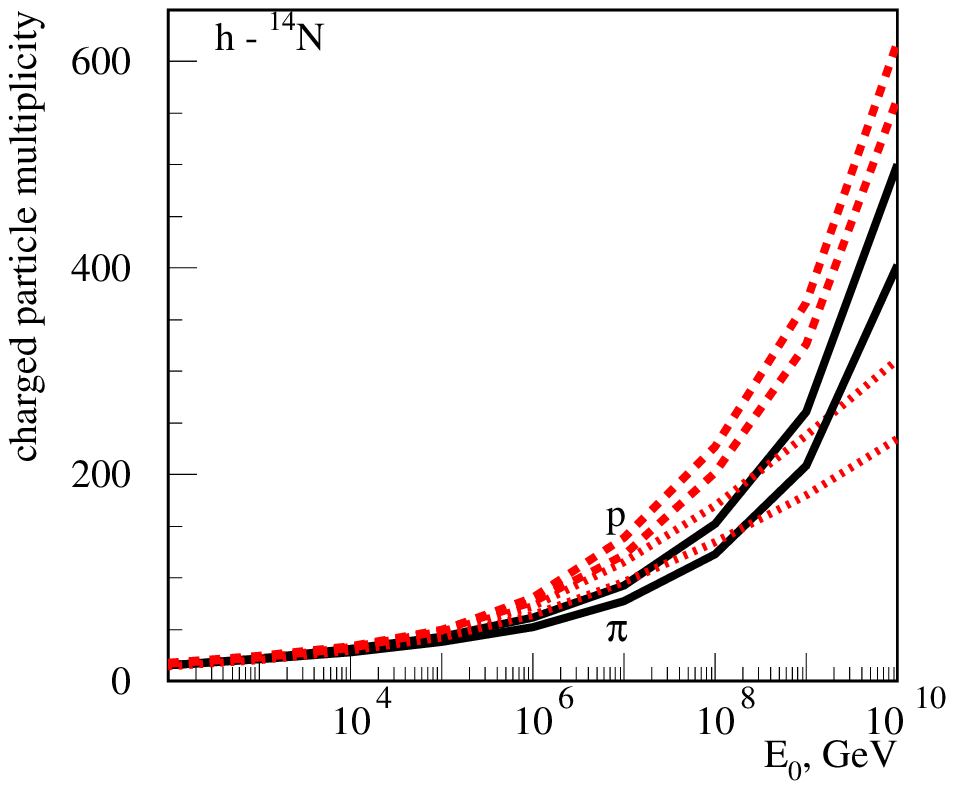}
\vspace*{-.9cm}
\end{center} 
\caption{Inelastic hadron-nitrogen cross section (left)
 and multiplicity of charged particles (right) for
  $p-^{14}\!N$ and $\pi -^{14}\!N$ collisions 
 as calculated with the QGSJET-II, QGSJET, and SIBYLL models --
    full, dashed, and dotted  curves correspondingly. \label{sig-air}} 
\end{figure} 
also increases with energy, reaching a factor of two at ultra-high energies, 
as shown in Fig.~\ref{sig-air}.
Main sources of these differences are the
treatment of semi-hard processes in the models,
 which depends on the chosen PDFs,
 and  the implementation of non-linear effects.
 The latter are not taken into account in  QGSJET, which employs
 flat (``pre-HERA'') parton distributions. 
 SIBYLL uses realistically steep PDFs, which leads to a faster
 energy increase of  $\sigma ^{\rm inel}_{\rm h-air}$.
   On the other hand, the faster multiplicity increase
in QGSJET is due to the semi-hard Pomeron scheme
employed, where  additional particle production comes from the
``soft pre-evolution'', i.e.~from hadronization of partons
of small $p_t$ ($p_t<Q_0$). It is noteworthy
that  cross sections in QGSJET-II  are rather similar
to the ones of  QGSJET -- see  Fig.~\ref{sig-air}(left),
despite the fact that the model uses realistic PDFs compatible with HERA
data. This is because the effect of new PDFs is compensated by 
non-linear screening corrections.
Such a compensation does not hold in SIBYLL, where such corrections
are introduced via a $Q_0(s)$ cutoff for hard processes and are 
neglected for the ``soft'' component. On the other hand,  non-linear
effects reduce particle production in QGSJET-II
compared to QGSJET,  bringing it closer to  
SIBYLL  -- see  Fig.~\ref{sig-air}(right). For  EAS
characteristics one obtains a rather large difference between QGSJET
and SIBYLL predictions at highest energies, 
which reaches in particular 30\% for the  muon number.
 In turn, for QGSJET-II  this difference is reduced to 10\% \cite{ost06}.

\section{Some remaining problems}
Among the most remarkable recent results are KASCADE studies of electron-muon
number correlations \cite{ant05}, which, apart from providing information
on the spectra and mass composition of cosmic rays in the knee region,
allowed one to quantify the discrepancies between present MC model predictions
and CR data. The analysis showed that both QGSJET and SIBYLL have problems
with the data interpretation. The corresponding discrepancies stay at the 
level of 10\% for the predicted $N_e$ and $N_{\mu}$ and arise for the two
models considered at different energy intervals. To reach a better agreement
with the data it is desirable to have a model which predicts slightly
deeper $X_{\max}$ compared to QGSJET, or, alternatively, reduces $N_{\mu}$,
moving towards SIBYLL predictions. In fact, both holds in QGSJET-II,
where non-linear effects sizably reduce particle production compared to
original QGSJET. Therefore, one may expect of it a somewhat better
performance here.

Apart from that,  present models seem to be unable to describe high
 multiplicity  muon bundles observed by LEP collaborations
\cite{tra05}. One could think of improving the situation via a significant
enhancement of multiplicity of hadronic interactions. On the other hand,
as the highest multiplicity bundles mainly originate from events with the
shower core inside the detector, one may expect a better performance from
models which predict a deeper $X_{\max}$, hence, a steeper muon
lateral distribution \cite{tra05}.

Finally, there exist a long-standing contradiction between ground-based
and fluorescence-based EAS detection techniques,  concerning both
obtained CR spectra
and  mass composition \cite{wat04}. While the composition issue is still
 difficult to address, the  energy reconstruction
methods, as discussed above,
 should not exhibit  significant model dependences. 
 Hence, further studies of systematics of corresponding experiments
 seem to be desirable.

\section{Future prospects}

During last decade one observed a convergence of EAS predictions of
contemporary cosmic ray interaction models. The remaining differences are still
significant, resulting in particular in sizable uncertainties of CR
composition studies. On the other hand, numerous new data, obtained with
cosmic rays or at colliders, open the way for model discrimination and for
their further development. Among the most important model uncertainties
are those connected to the energy extrapolation of inelastic cross sections
and of the inelasticity of hadron-air interactions. The new measurement
of $\sigma ^{\rm inel}_{\rm h-air}$ by the HIRES collaboration \cite{bel05}
allows us to reduce the former and points towards smaller cross sections
than those of present models. Future LHC measurements of 
$\sigma ^{\rm tot}_{\rm pp}$ will give us  possibilities
 to calibrate reliably corresponding
model procedures. On the other hand, RHIC collider provided valuable
data on baryon ``stopping'' in central nucleus-nucleus collisions \cite{rhic}.
When repeated for different centrality selections and for various 
combinations of projectile and target nuclei, those will allow us to fix
reliably model predictions for the inelasticity of hadron-nucleus interactions.

On the theoretical side, of great importance is  the
pQCD treatment of non-linear interaction effects, in particular, 
concerning particle production in the transition region close to the
parton saturation regime. Here, one may expect further progress to come
from the ``quasi-classical QCD'' approach \cite{mcl94}.

\bigskip
{\small The author would like to acknowledge fruitful discussions with R. Engel.}

\end {document}